\documentclass[a4paper,twoside]{article}
\usepackage[T1]{fontenc}
\usepackage{graphicx}
\usepackage{epsfig}
\usepackage{pslatex}
\usepackage[bottom]{footmisc}
\usepackage{apalike}
\usepackage{algorithm2e}

\usepackage{tikz}
\usetikzlibrary{arrows.meta, positioning, calc}
\usepackage{comment}
\usepackage{tabularx}
\usepackage{booktabs}
\usepackage{multirow}
\usepackage{float}
\usepackage{xcolor}
\usepackage{listings}
\usepackage{hyperref}
\usepackage{amsmath}
\usepackage{subcaption}
\usepackage{color}
\usepackage{etoolbox}
\usepackage{SCITEPRESS}

\newif\ifblindreview
\blindreviewfalse
\ifblindreview
  \newcommand{\BPTakiCite}{Anon26}
  
\else
  \newcommand{\BPTakiCite}{AsYaKa26a}
  
\fi

\definecolor{codegreen}{rgb}{0,0.6,0}
\definecolor{codegray}{rgb}{0.5,0.5,0.5}
\definecolor{codepurple}{rgb}{0.58,0,0.82}
\definecolor{backcolour}{rgb}{0.98,0.98,0.96}
\definecolor{deepblue}{rgb}{0,0,0.5}

\hypersetup{
    colorlinks=true,
    allcolors=deepblue
}

\lstdefinelanguage{BPPy}
{morekeywords={interrupt, block, request, waitFor, Block, Request, mustFinish, localReward, yield, while, def, for, in, range, from, import, return, if, elif, else, lambda, sum, filter, class, super},
sensitive=false,
morecomment=[l]{\#},
morecomment=[s]{/*}{*/},
morestring=[b]",
morestring=[b]',
}

\lstdefinestyle{mystyle}{
    language=BPPy,
    backgroundcolor=\color{backcolour},   
    commentstyle=\color{codegreen},
    keywordstyle=\color{blue},
    keywordstyle=[5]\ttfamily\scriptsize
    numberstyle=\tiny\color{codegray},
    stringstyle=\color{codepurple},
    basicstyle=\ttfamily\small,
    breakatwhitespace=false,         
    breaklines=true,                 
    captionpos=b,                    
    columns=fullflexible,
    keepspaces=true,                 
    numbers=none,                    
    numbersep=5pt,                  
    showspaces=false,                
    showstringspaces=false,
    showtabs=false,
    tabsize=2,
    aboveskip=4pt,
    belowskip=2pt
}

\newcommand{\li}[2][]{\lstinline[basicstyle=\ttfamily\small,breaklines=false,#1]{#2}}

\renewcommand\j[1]{\textsc{#1}}

\begin{document}
\title{From Domain Knowledge to Composable Reactive Code: LLM-Assisted Development with Scenario-Based Programming}
\ifblindreview
\author{\authorname{Anonymous Author(s)}
\affiliation{Affiliation withheld for review}
\email{}
}
\else
\author{\authorname{Adiel Ashrov\sup{1}\orcidAuthor{0000-0003-4510-5335}, Tom Yaacov\sup{2}\orcidAuthor{0000-0002-0565-6506} and Guy Katz\sup{1}\orcidAuthor{0000-0001-5292-801X}}
\affiliation{\sup{1}The Hebrew University of Jerusalem, Jerusalem, Israel}
\affiliation{\sup{2}King's College London, London, United Kingdom}
\email{\{adiel.ashrov,g.katz\}@mail.huji.ac.il, tom.yaacov@kcl.ac.uk}
}
\fi

\abstract{Reactive systems must continuously respond to events and coordinate interacting behaviors, making code generation challenging: reliability depends not only on individual components but on how they compose at runtime. We investigate \emph{Scenario-Based Programming (SBP)} in two complementary roles: as a structured medium for communicating domain knowledge to a \emph{Large Language Model (LLM)}, and as the target paradigm for generated extensions. We implement the \emph{Taki} card game in SBP and ask an LLM to generate Python players using progressively richer SBP-cast domain knowledge. Separately, we ask the LLM to generate strategy scenarios that compose directly with the existing implementation. In controlled simulations, all generated strategies that we evaluated complete their runs without observed runtime errors, deadlocks, or draws. The generated SBP strategy also composes with an additional hand-authored scenario without observed execution failures. Richer SBP-cast domain knowledge is associated with stronger generated Python strategies: their win rate against a fixed random SBP baseline increases from 59.4\% to 62.5\%. The directly generated SBP strategy wins 64.9\% against the same baseline and 51.2\%--52.3\% head-to-head against the generated Python strategies. These results suggest that SBP can serve both as a medium for communicating domain knowledge and as a composable target for reactive-system extensions.}

\onecolumn \maketitle \normalsize \setcounter{footnote}{0} \vfill

\section{\uppercase{Introduction}}

\emph{Large Language Models (LLMs)} increasingly assist 
software developers with code-generation 
tasks~\cite{ChTwJuYuDeKaEdBuJoBr21,LiXiWaZh23,YaElWe24,HaKaMaSz25}. 
However, generating reliable code for complex reactive systems 
remains challenging~\cite{LiXiWaZh23,BeKa26}. In such systems, 
correctness depends not only on the behavior of individual 
components in isolation, but also on how they coordinate over 
time. A generated component may therefore appear correct in 
isolation while violating system-level assumptions when 
integrated with existing code, for example, by enabling invalid 
event sequences, blocking required actions, or disrupting the 
coordination among existing behaviors.
This distinguishes two tasks. In isolated code generation, a model
produces code from a self-contained specification, and the result
can be judged on its own. In compositional extension, the model adds
code to an existing system, and the result must also work with what
is already there. Recent studies of repository-level code generation
show that using the surrounding codebase remains
challenging~\cite{LiLiZhLiLiZhWaLiFaWaDiZhZhDoJiLiHuLiGuYa24,HaNgBu25}; for example, some models
reimplement existing dependencies instead of using
them~\cite{HaNgBu25}. Reactive systems add a further challenge: new
code must also respect how existing components coordinate at
runtime. This motivates development approaches that expose these
constraints to the LLM in a form that supports compositional
extension.

One promising approach to addressing this challenge is 
\emph{Scenario-Based Programming (SBP)}~\cite{HaMaWe12}, which 
is a generalization of \emph{Live Sequence Charts (LSC)}~\cite{DaHa01}. 
In SBP, system behavior emerges 
from independent \emph{b-threads}, each typically capturing a 
single concern, which a central execution engine weaves 
together by selecting requested events that are not blocked. 
This architecture supports modularity, since each 
b-thread expresses a requirement independently, and composability, 
since new concerns can often be addressed by 
introducing additional b-threads without modifying existing 
ones. Prior work suggests that this decomposition aligns with 
how people naturally describe system 
requirements~\cite{GoMaMe12}, motivating the hypothesis that 
SBP may ease both LLM-assisted implementation and the 
communication of domain knowledge to an LLM through a prompt. 
Despite these apparent affinities, the synergy between SBP and 
LLM-assisted code generation has received limited empirical 
evaluation, particularly for incremental 
extensions to an existing reactive system.

In this paper, we investigate the combination of SBP and LLM-assisted code generation in two complementary roles: as a structured representation through which domain knowledge is communicated to the LLM, and as the target paradigm in which generated extensions are implemented. As a testbed, we implement \emph{Taki}~\cite{ShGa26a}---a complex card game involving action cards, multi-step sequences, and asymmetric player information---using BPpy~\cite{YaWeAsKaZi25}. We ask a target LLM to generate Python players under progressively richer contexts containing domain knowledge cast through SBP, and separately to generate strategy b-threads that compose directly with the existing SBP implementation. We evaluate the resulting strategies through repeated controlled simulations. Specifically, we study whether the generated Python and SBP strategies execute and integrate without observed failures (\textbf{RQ1}), whether the generated SBP strategy achieves competitive strategic performance relative to the generated Python players (\textbf{RQ2}), and whether richer domain knowledge cast through SBP improves LLM-generated Python players and, when cast as strategy b-threads, improves performance over a random baseline (\textbf{RQ3}).

Together, these two roles position SBP as a knowledge representation medium for cooperation between software engineers and AI models during AI-assisted software development. Engineers can express domain rules and strategies as b-threads and provide them to a model, and the model can return extensions as separate b-threads that compose with the existing ones. In this work, we evaluate the model's side of this exchange; how engineers write, understand, and review such b-threads remains to be studied.

We evaluate the generated strategies through 10{,}000-game
pairwise simulations with balanced starting positions and a fixed seed
schedule, reporting overall and position-conditioned win rates together
with Wilson score confidence intervals (Section~\ref{sec:experimental-setup}).
The results, detailed in Section~\ref{sec:results}, support an affirmative
answer to all three research questions: the generated strategies execute
and integrate without observed failures (\textbf{RQ1}); the directly
generated SBP strategy remains competitive with the generated Python
strategies (\textbf{RQ2}); and the strongest generated
Python strategy was obtained with the richest SBP-cast context
(\textbf{RQ3}).

The rest of the paper is organized as follows.
Section~\ref{sec:sbp-and-bppy} introduces SBP and the BPpy framework.
Section~\ref{sec:taki-a-complex-reactive-system} presents the Taki
testbed, its SBP implementation, and the hand-authored strategy
refinements used in the study. Section~\ref{sec:experimental-setup}
describes the research questions, strategy-generation pipeline,
experimental configurations, evaluation criteria, and simulation
protocol. Section~\ref{sec:results} reports the experimental results,
and Section~\ref{sec:discussion} discusses their implications and
threats to validity. Section~\ref{sec:related-work} reviews related
work, and Section~\ref{sec:conclusions} concludes and outlines
directions for future work.

\section{\uppercase{Scenario-Based Programming and BPpy}}
\label{sec:sbp-and-bppy}

We begin by introducing the key concepts of scenario-based programming (SBP). For a grounded discussion, we illustrate these concepts using the BPpy package~\cite{YaWeAsKaZi25}, a framework for SBP in Python.

In SBP, system behavior is specified through scenarios named \emph{b-threads}. These are sequential threads of execution, each specifying behavior that the system should include or avoid. B-threads are usually independent and self-contained, with each one focusing on a particular aspect of the system, often corresponding to an individual requirement. At runtime, an application-agnostic execution mechanism coordinates the b-threads and combines their individual specifications into a unified system behavior. This coordination follows the synchronization protocol of behavioral programming~\cite{HaMaWe10}. Before every event generated by the b-program is selected, each b-thread submits a synchronization statement declaring which events it \emph{requests}, which events it \emph{waits for} (without actively requesting), and which events it \emph{blocks} from occurring. After submitting its statement, the b-thread is paused. When all b-threads have submitted their statements, the b-program reaches a \emph{synchronization point}. An event arbiter then selects one event that is requested by at least one b-thread and is not blocked by any b-thread. All b-threads that requested or waited for the selected event are resumed, while the remaining b-threads stay paused and retain their statements for the next synchronization point. This synchronization and event-selection cycle continues throughout the execution of the b-program.

In addition to request, waitFor, and block, the BPpy
event-selection strategy used in this work associates requested events with 
numeric priorities. Lower values denote stronger selection 
preference. At each synchronization point, the arbiter 
removes blocked events, keeps the requested events with the 
lowest priority value, and selects uniformly at random 
among ties. We use this mechanism to encode preferences 
among simultaneously available actions, as described in
Sections~\ref{sec:sbp-taki-implementation} 
and~\ref{sec:baselines}. The formal semantics of SBP are available in the supplementary material~\cite{\BPTakiCite}.

To illustrate these concepts, we begin with a simple example b-program implemented using BPpy. The example adapts a published sample b-program~\cite{HaMaWe12}, in which a system controls the mixing of hot and cold water from two separate taps. \autoref{lst:add-hot-cold} presents two b-threads, \li{add_hot} and \li{add_cold}, each of which requests three events representing the pouring of a small amount of hot or cold water. In BPpy, b-threads are implemented as Python generators, i.e., functions that can suspend their execution and return data to the caller using the \li{yield} idiom, and later resume execution when invoked through the \li{send} method. Synchronization statements are represented by instances of the \li{sync} class, which specify events or sets of events through the \li{request}, \li{block}, and \li{waitFor} arguments. The BPpy execution mechanism initially invokes each b-thread generator independently and waits for it to yield a synchronization statement. After collecting statements from all b-threads, the mechanism selects an event and continues the execution of the b-program according to the synchronization protocol described above.

\begin{lstlisting}[
label={lst:add-hot-cold},
caption={The \li{add_hot} and \li{add_cold} b-threads. Each requests its corresponding event three times.}
]
@bp.thread
def add_hot():
    for i in range(3):
        yield bp.sync(
            request=bp.BEvent("HOT"))
@bp.thread
def add_cold():
    for i in range(3):
        yield bp.sync(
            request=bp.BEvent("COLD"))
\end{lstlisting}

In contrast to many programming paradigms, SBP does not require developers to define a single predetermined system behavior. Rather, the system may follow any behavior that is consistent with the constraints imposed by all specified b-threads. For example, a b-program containing only the two b-threads in \autoref{lst:add-hot-cold} places no restrictions on the order in which hot and cold water are poured. Its execution may produce any sequence containing exactly three occurrences of the \li{HOT} event and three occurrences of the \li{COLD} event, such as \li{COLD, HOT, HOT, COLD, HOT, COLD} or \li{COLD, COLD, COLD, HOT, HOT, HOT}.

To further illustrate, suppose that, after the initial system has been in use for some time, a safety issue is identified and a new requirement is introduced: two \li{HOT} events should not occur consecutively. One possible solution is to modify the existing \li{add_hot} b-thread by incorporating additional conditions and synchronization statements. However, SBP encourages maintaining the alignment between individual b-threads and the requirements they represent by introducing a separate b-thread for the new requirement. This supports an incremental and modular development process, in which behaviors can be added or removed independently without modifying existing b-threads. Accordingly, \autoref{lst:control} introduces the \li{control} b-thread, which repeatedly waits for a \li{HOT} event and, once such an event occurs, blocks the next \li{HOT} event while waiting for any subsequent event through the \li{All} event set.

This example previews how we use SBP in the rest of the paper. Just
as \li{control} blocks an event, Taki's rule b-threads block illegal
actions (Section~\ref{sec:sbp-taki-implementation}). And just as it
adds a requirement without changing \li{add_hot}, strategy b-threads
extend the game without changing its rule b-threads, by blocking
events or re-requesting them with stronger priorities
(Section~\ref{sec:hand-authored-sbp-strategies}).

SBP has also been implemented in other languages, including Live
Sequence Charts (LSC)~\cite{DaHa01,HaMa03},
JavaScript~\cite{BaWeSh18}, C++~\cite{HaKa14}, and Java~\cite{HaMaWe10}, as well as in
dedicated frameworks~\cite{GrGrKaMa16}; and it has been applied across domains such as a reactive IoT building~\cite{ElMaStWe18} and a nano-satellite~\cite{BaElSaWe19}; the concepts presented here may apply to these settings as well.

\begin{lstlisting}[
label={lst:control},
caption={The \li{control} b-thread, which prevents the \li{HOT} event from occurring consecutively.}
]
@bp.thread
def control():
    while True:
        yield bp.sync(
            waitFor=bp.BEvent("HOT"))
        yield bp.sync(
            waitFor=bp.All(),
            block=bp.BEvent("HOT"))
\end{lstlisting}

\section{\uppercase{TAKI as a Complex Reactive System}}
\label{sec:taki-a-complex-reactive-system}

This section first introduces the rules of the Taki variant that we
study. It then describes its SBP implementation
(Section~\ref{sec:sbp-taki-implementation}), shows how hand-authored
strategies extend that implementation
(Section~\ref{sec:hand-authored-sbp-strategies}), and explains why the
game tests composition
(Section~\ref{sec:taki-stress-test-composition}).

\emph{Taki} is a shedding-type card game in which the first player 
to discard all cards from their hand wins~\cite{ShGa26a}. At the 
beginning of a game, each player receives eight cards, the remaining 
cards form the \emph{draw pile}, and one card is placed face-up to 
start the \emph{discard pile}. The top card of this pile, referred 
to as the \emph{leading card}, determines which cards may be played 
next. On each turn, the active player may discard a card that 
matches the color, number, or type of the leading card. Certain 
action cards may also be played according to their own rules. If the 
player has no legal card, the player draws one card and the turn passes 
to the next player. Thus, even before considering action cards, the 
set of legal moves changes after every discarded card and depends on 
both the current leading card and the active player's hand. A visual 
overview of the gameplay is available in the accompanying video 
tutorial~\cite{ShGa26b}.

\begin{figure*}[t]
    \centering

    \begin{subfigure}[t]{0.37\textwidth}
        \centering
        \includegraphics[scale=0.056]{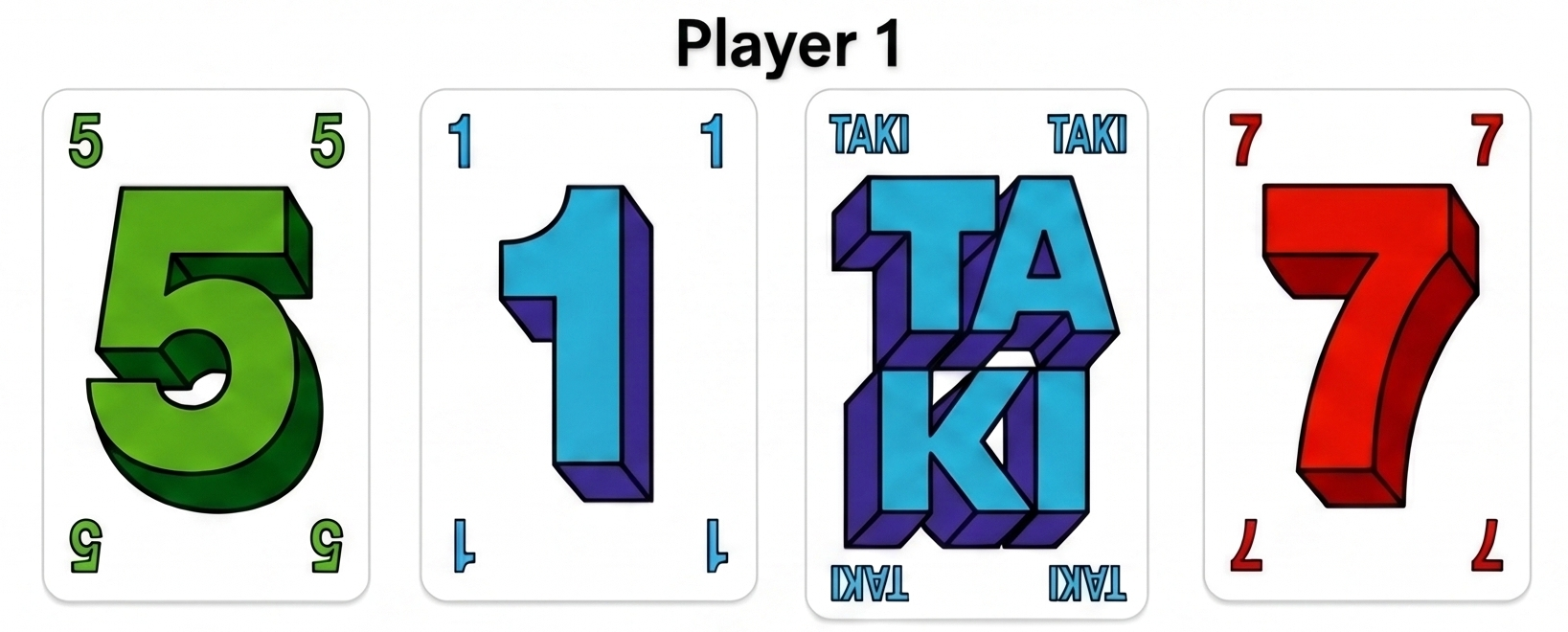}
    \end{subfigure}
    \hfill
    \begin{subfigure}[t]{0.23\textwidth}
        \centering
        \includegraphics[scale=0.105]{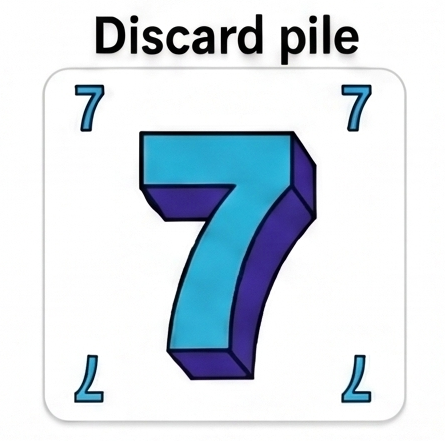}
    \end{subfigure}
    \hfill
    \begin{subfigure}[t]{0.37\textwidth}
        \centering
        \includegraphics[scale=0.056]{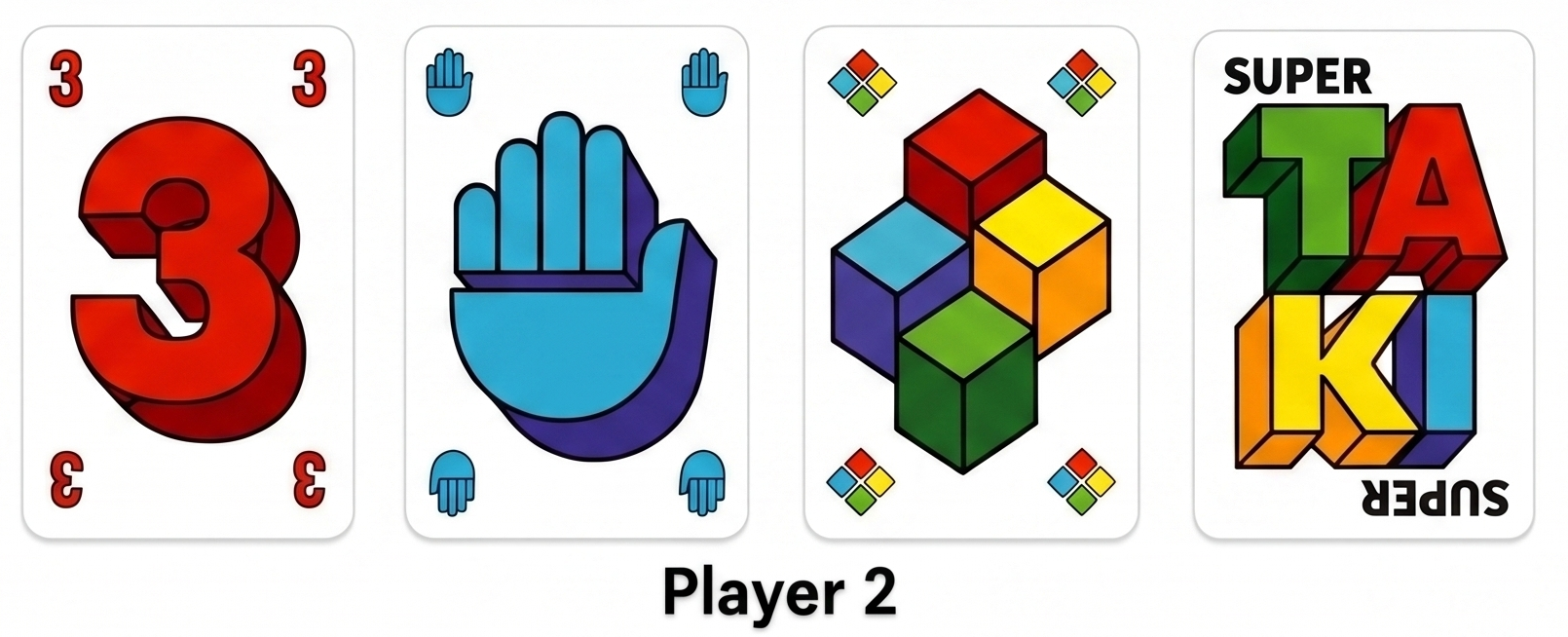}
    \end{subfigure}

    \caption{Illustration of a two-player Taki game state, with card 
    (\j{7}, \j{Blue}) as the leading card in the discard pile. 
    If Player~1 is active, the legal moves are 
    (\j{1}, \j{Blue}) and (\j{TAKI}, \j{Blue}), 
    which match its color, and (\j{7}, \j{Red}), which matches 
    its number; (\j{5}, \j{Green}) is not playable. 
    If Player~2 is active, the legal moves are 
    (\j{Stop}, \j{Blue}), which matches the leading color, 
    and the wild \j{Change-Color} and \j{Super TAKI} cards; 
    (\j{3}, \j{Red}) is not playable.}
    \label{fig:taki-game-illustration}
\end{figure*}

The official deck contains numbered cards in several colors 
together with action cards that modify the normal turn structure. We 
implement a two-player variant restricted to numbered cards and four 
action-card types. A \emph{Stop} card skips the opponent's next 
turn; consequently, in the two-player setting, the player who uses 
it immediately plays again. A \emph{TAKI} card opens a sequence in 
which the active player may discard multiple cards of the TAKI 
card's color during the same turn, until the player explicitly 
closes the sequence. A \emph{Super TAKI} card is a wild TAKI card 
that may be played independently of the leading card and opens 
a sequence by inheriting the current leading color. A 
\emph{change-color} card may also be played independently of the 
leading color and allows the active player to choose the color that 
constrains subsequent play. These cards introduce state that 
persists beyond the single action that creates it: the system must 
track whether a TAKI sequence is open, which color governs that 
sequence, whether the opponent's turn is skipped, and which color is 
selected after a change-color card. 
Figure~\ref{fig:taki-game-illustration} illustrates the 
ordinary matching rule and the legal action cards in 
a representative game state. The remaining action cards defined in 
the official ruleset, as well as multi-player play, are outside the 
scope of this implementation.

\subsection{The SBP Taki Implementation}
\label{sec:sbp-taki-implementation}

We implement the two-player Taki game as a b-program 
whose b-threads
separate action generation, rule enforcement, and strategic
preference. One instance of \li{player\_behavior} is created for
each player and maintains that player's private hand. These instances
request the actions currently available from their local perspective,
including playing each held card, drawing a card, selecting a color,
and closing a TAKI sequence. Separate rule-enforcement b-threads
constrain these requests. In particular, \li{enforce\_turns}
blocks actions of the non-active player, whereas
\li{enforce\_card\_placement} blocks card plays that violate the
current placement constraint. Additional b-threads manage dealing, game termination, and execution monitoring. The
complete executable implementation, including omitted helper methods
and initialization protocols, is available in the project
repository~\cite{\BPTakiCite}.

The b-threads coordinate through player-specific events such as
\li{P0:Play(card)}, \li{P0:DrawCard}, and
\li{P0:ClosedTaki}, together with global control events such as
\li{NextTurn} and \li{DonePostAction}. Each requested event
carries a numeric priority, where a lower value denotes a stronger
selection preference. At each synchronization point, the event-selection
mechanism first collects the events requested by all b-threads and
removes every event blocked by at least one b-thread. It then retains
only requests having the lowest priority value among the remaining
events and selects uniformly at random among them. Strategy b-threads may re-request selected actions with stronger priorities, thereby biasing the arbiter toward preferred legal moves without modifying the rule-enforcement b-threads.

Listing~\ref{lst:player-behavior} presents the b-thread that exposes
one player's possible actions to the event-selection mechanism. The
b-thread maintains only the player's local hand; it does not determine
whether a requested card is currently legal or whether it is the
player's turn. Instead, it requests one \li{Play(card)} event for
each held card and relies on independent rule b-threads to block
inadmissible actions. This separation allows the same player behavior
to operate with different rule and strategy b-threads.

\begin{lstlisting}[
label={lst:player-behavior},
caption={Core of the \li{player\_behavior} b-thread. It requests
every held card or a lower-priority draw. The omitted helper
\li{update\_hand} updates the hand and handles TAKI and change-color
continuations. A more detailed version appears in the supplementary
material~\cite{\BPTakiCite}.},
float=!tbp
]
@bp.thread
def player_behavior(index, initial_hand):
  hand = list(initial_hand)
  while hand:
    event = yield bp.sync(request=
      [play_event(index, c, priority=10.0)
       for c in hand] +
      [draw_event(index, priority=20.0)])
    yield from update_hand(index, hand, event)
    if hand:
      yield bp.sync(request=BPEvent(
          "NextTurn", priority=10.0))
  yield bp.sync(
      request=no_more_cards_event(index))
\end{lstlisting}

Listing~\ref{lst:enforce-turns} presents the turn-order b-thread, which tracks the active player, blocks the opponent's actions, and handles the special control flow introduced by Stop and TAKI-family cards. After a Stop card or a closed TAKI sequence, it requests \li{DonePostAction} while blocking all player actions. Other b-threads, such as the player and strategy b-threads, wait for this event before continuing, which ensures that no player action occurs before the current one is complete.

\begin{lstlisting}[
label={lst:enforce-turns},
float=!tbp,
caption={Core of the \li{enforce\_turns} b-thread for the
two-player variant. It blocks the opponent's actions, lets the same
player act again after a Stop card, and keeps the opponent blocked
during a TAKI sequence. Helper definitions are omitted; a more
detailed version appears in the supplementary
material~\cite{\BPTakiCite}.}
]
@bp.thread
def enforce_turns(player=0):
  next_player = 1 - player
  while True:
    opp = other_player_action_events(player)
    event = yield bp.sync(
        waitFor=[BPEvent("NextTurn"),
          player_play_event_set(player)],
        block=opp)
    if is_next_turn(event):
      player = next_player
      next_player = 1 - player
    elif is_stop(event):
      next_player = player
      yield bp.sync(
          request=BPEvent(
            "DonePostAction", priority=10.0),
          block=all_player_action_event_set())
    elif is_taki_family(event):
      yield bp.sync(
          waitFor=player_event(
              player, "ClosedTaki"),
          block=opp)
      yield bp.sync(
          request=BPEvent(
            "DonePostAction", priority=10.0),
          block=all_player_action_event_set())
\end{lstlisting}

For validation, we implemented three monitoring b-threads that observe generated execution traces: one checks that every player action belongs to the active player, a second detects deadlock when no requested and unblocked event is available, and a third bounds the total number of player events in a game, recording the run as a draw when this bound is exceeded.

\subsection{Hand-Authored SBP Strategy Refinements}
\label{sec:hand-authored-sbp-strategies}

To illustrate how strategic knowledge can be added to the SBP Taki
implementation, we implemented three hand-authored strategy refinements.
These refinements are designed by the researchers as simple heuristics
for improving the basic random SBP player. They are not claimed to be
optimal Taki strategies; rather, they provide controlled examples of how
strategy-level knowledge can be expressed as modular b-threads that
compose with the rule-enforcement b-threads described above.

The first refinement, \j{TAKI-family-preference}, assigns the same
stronger selection preference to all TAKI-family cards, including both
regular TAKI and Super TAKI. The motivation is that TAKI-family cards
can open multi-card sequences, and therefore may allow the player to
reduce their hand more quickly than ordinary single-card moves. The implementation of this strategy is available in the supplementary material~\cite{\BPTakiCite}.

The second refinement, \j{TAKI-over-Super-TAKI}, introduces a more fine-grained priority hierarchy among TAKI-family cards. Regular TAKI cards receive the strongest selection preference, Super TAKI cards receive an intermediate selection preference, and other cards keep their default priority. Thus, the strategy still prefers TAKI-family cards over ordinary cards, but also prefers opening a regular TAKI over Super TAKI when both are available. Listing~\ref{lst:prefer-taki-over-super-taki} shows the core of this strategy b-thread; helper definitions are omitted.

The third refinement, \j{Conserve-Super-TAKI}, combines
\j{TAKI-over-Super-TAKI} with an additional b-thread that blocks Super
TAKI while a regular TAKI sequence is already open. The motivation is
to conserve Super TAKI for later states in which its wild-card behavior
may be more useful, instead of spending it inside a sequence that is
already constrained by the color of a regular TAKI. The implementation of this strategy is available in the supplementary material~\cite{\BPTakiCite}.

\begin{lstlisting}[
label={lst:prefer-taki-over-super-taki},
float=!tbp,
caption={Core of the
\texttt{prefer\_\allowbreak{}taki\_\allowbreak{}over\_\allowbreak{}super\_\allowbreak{}taki\_\allowbreak{}strategy}
b-thread. It re-requests the held cards, giving regular TAKI
priority~4.0, Super TAKI~6.0, and other cards the default~10.0.
After an action card, it waits for \li{DonePostAction} before
requesting cards again. The omitted helper
\li{update\_strategy\_hand} updates the hand and applies the same
priorities during a TAKI sequence. A more detailed version appears in
the supplementary material~\cite{\BPTakiCite}.}
]
@bp.thread
def prefer_taki_over_super_taki_strategy(
    index, initial_hand):
  hand = list(initial_hand)
  def prio(card):
    return (4.0 if is_regular_taki(card)
        else 6.0 if is_super_taki(card)
        else 10.0)
  while hand:
    event = yield bp.sync(
      request=[
        play_event(index, c, priority=prio(c))
        for c in hand],
      waitFor=draw_event(index))
    yield from update_strategy_hand(
        index, hand, event, prio)
    if is_action_card(event):
      yield bp.sync(
          waitFor=BPEvent("DonePostAction"))
\end{lstlisting}

Importantly, all three refinements leave the rule-enforcement b-threads
unchanged. The first two refinements influence event selection by
assigning lower numeric priority values to preferred events, while the
third adds a temporary blocking constraint. This illustrates the
compositional style of SBP: strategic knowledge can be added as small
behavioral modules that interact with the existing game implementation
through the shared event-selection mechanism.

\subsection{TAKI as a Stress Test for Composition}
\label{sec:taki-stress-test-composition}

Taki combines several sources of reactive complexity, including action cards that modify the normal turn structure, multi-event interactions such as TAKI sequences, and asymmetric player information. In our implementation, these concerns are captured by independent b-threads that interact to yield complex system behavior, making Taki a useful testbed for composition in reactive systems. Although Taki is turn-based, turns do not strictly alternate, nor does each consist of a single action. In our two-player variant, a Stop card lets the same player act again, and a TAKI-family card opens a multi-event procedure during which the opponent remains blocked until the sequence is closed, while rule b-threads constrain the cards played within it. However, such procedures are initiated by the active player and are not preempted by the opponent. Our testbed therefore does not exercise asynchronous interrupts, such as external events that preempt an ongoing procedure, and we do not claim that our results extend to systems in which such interrupts are essential.

In this setting, generated strategies must compose correctly with existing rule-enforcement b-threads rather than implement the game logic independently. This makes Taki a suitable benchmark for evaluating whether LLM-generated behavioral extensions integrate
into an existing reactive system without observed execution failures, a question revisited in Section~\ref{sec:results}.

\section{\uppercase{Experimental Setup}}
\label{sec:experimental-setup}

We next describe the research questions and metrics, the generation
pipeline, the prompt configurations, the baselines, and the simulation
protocol.

\subsection{Research Questions and Metrics}
\label{sec:research-questions}
\label{sec:evaluation-dimensions}

\paragraph{RQ1 --- Executability and Integration.}
Can LLM-generated strategies, whether Python players or SBP strategy
b-threads, be executed and integrated into the Taki simulation framework
without observed execution failures? We monitor all runs for runtime
errors, deadlocks, and draws. A draw denotes a run terminated by the
livelock detector (Section~\ref{sec:sbp-taki-implementation}) rather
than a rule-based Taki outcome. Since it can also arise from a legal but
non-terminating interaction between well-formed strategies, we treat it
as a warning signal rather than a confirmed defect. For \j{SBP-Native},
we also run self-play with and without an additional hand-authored
b-thread, in both player positions
(Section~\ref{sec:executability}).

\paragraph{RQ2 --- Strategic Performance.}
Can LLM-generated SBP strategy b-threads achieve competitive performance
compared with LLM-generated Python players under the same conditions?
We play \j{SBP-Native} head-to-head against each LLM-generated Python
strategy. The primary metric is win rate. Since Taki exhibits a
first-player advantage, we report it overall and by starting position,
with 95\% Wilson score confidence intervals. We compare Python strategies'
win rates against \j{SBP-Random} using exact McNemar tests~\cite{Mc47},
pairing games by seed, and test head-to-head win rates against 50\% using
exact binomial tests. Tests are two-sided, with Holm
adjustment~\cite{Ho79} within each group of three. All tests concern
overall win rates.

\paragraph{RQ3 --- Domain Knowledge.}
Does richer domain knowledge, cast through SBP, improve LLM-generated
Python players, and does the same knowledge, cast as SBP strategy
b-threads, improve performance over the same random baseline? We compare
\j{Py-Base}, \j{Py-SBP-Core}, and \j{Py-SBP-Strategies} against the same
\j{SBP-Random} opponent. Since \j{SBP-Native} changes the generated
artifact and has only a single configuration, we report it as a separate
reference point rather than as another step in the Python progression.

\subsection{Experimental Pipeline}
\label{sec:experimental-pipeline}

\begin{figure*}[t]
\centering
\begin{tikzpicture}[
  node distance=6mm and 4mm,
  box/.style={draw, thick, rounded corners=2pt, minimum height=0.9cm,
              minimum width=1.9cm, align=center, font=\footnotesize,
              fill=gray!8},
  numcirc/.style={circle, draw, thick, fill=white, minimum size=4mm,
                  font=\scriptsize\bfseries, inner sep=0pt},
  arr/.style={-{Latex[length=2mm, width=1.5mm]}, thick}
]

\node[box] (n1) {Context\\Definition};
\node[box, right=14mm of n1] (n2) {Prompt\\Design};
\node[box, right=14mm of n2] (n3) {LLM\\Input};
\node[box, right=of n3] (n4) {Generate\\Strategy};
\node[box, right=of n4] (n5) {Validation \&\\Evaluation};

\node[numcirc, above=1mm of n1.north] {1};
\node[numcirc, above=1mm of n2.north] {2};
\node[numcirc, above=1mm of n3.north] {3};
\node[numcirc, above=1mm of n4.north] {4};
\node[numcirc, above=1mm of n5.north] {5};

\draw[arr] (n1) -- (n2) node[midway, above, font=\footnotesize, yshift=0.5mm] {Context};
\draw[arr] (n2) -- (n3) node[midway, above, font=\footnotesize, yshift=0.5mm] {Prompt};
\draw[arr] (n3) -- (n4);
\draw[arr] (n4) -- (n5);

\coordinate (c1) at ($(n1.south)+(0,-6mm)$);
\coordinate (c2) at (n3.south |- c1);
\draw[arr] (n1.south) -- (c1) -- (c2) -- (n3.south);
\node[font=\footnotesize, above] at ($(c1)!0.5!(c2)$) {Context};

\end{tikzpicture}
\caption{The experimental pipeline used for strategy generation and evaluation.}
\label{fig:experimental-pipeline}
\end{figure*}
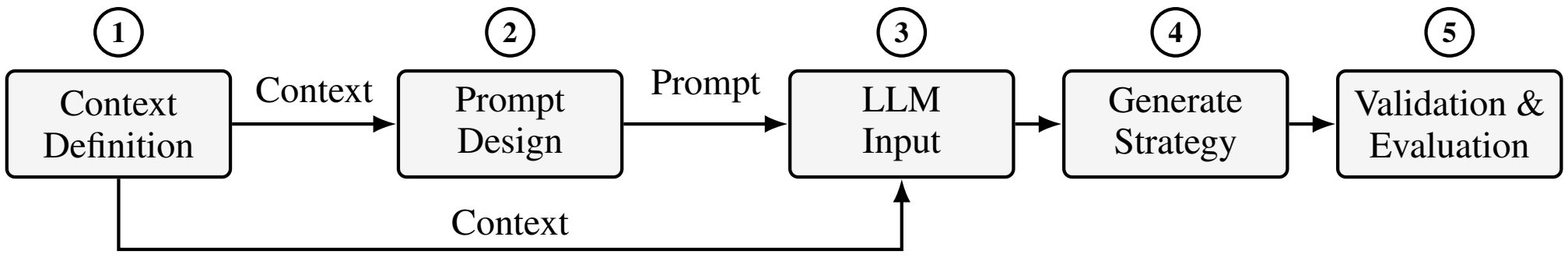

All experiments follow the same pipeline, shown in Figure~\ref{fig:experimental-pipeline}. 
For each configuration, we first define the context by selecting the available source files and supporting documentation. We then manually write a prompt, which is refined with the assistance of \emph{Claude Sonnet 4.6}~\cite{An26}. Claude is used only for this refinement, so the target LLM does not author its own instructions.
The context and prompt are then provided as input to
\emph{Codex}~\cite{Op25}, using \emph{GPT-5.4}~\cite{Op26}
as the target LLM, which generates the strategy code.
During generation, the target LLM may inspect
the source files, run tests, observe runtime errors, and revise its
code, but it receives no additional human guidance, iterative prompting,
or corrective feedback. The resulting strategy is then validated and
evaluated according to the criteria described in Section~\ref{sec:evaluation-dimensions}.
The pipeline remains fixed across configurations; only the context and its 
corresponding prompt vary, thereby determining the target implementation
form and the extent to which SBP-cast domain knowledge is
available to the model. 

\subsection{Experimental Configurations}
\label{sec:experimental-configurations}

Table~\ref{tab:prompts-configurations} summarizes the four
configurations. All include the official Taki rulebook and the same
task: generate a strategy for one Taki player that selects or
prioritizes actions in order to win. They form an incremental
progression. In \j{Py-Base}, the model receives only the rulebook and
the Python player interface, which maps a player-facing observation of
the game state to a chosen action. \j{Py-SBP-Core} adds the BPpy-based
SBP Taki implementation, which shows how the rules are encoded in SBP.
\j{Py-SBP-Strategies} further adds the hand-authored strategy b-threads
of Section~\ref{sec:hand-authored-sbp-strategies} as examples of
behaviorally expressed strategic knowledge. \j{SBP-Native} changes the
target artifact: given the SBP implementation, the BPpy API and idioms,
and the same examples, the model generates new strategy b-threads that
compose directly with the existing program. We use \emph{native} to
convey that this artifact is expressed directly in SBP, whereas the
first three configurations produce Python players. All configurations
and prompts are available in the project
repository.\footnote{\url{https://github.com/adielashrov/taki-llm-experiments}}

\begin{table*}[t]
\centering
\caption{Prompt configurations. All include the official rulebook.
Input strategy examples are context for the target LLM, not the
evaluated strategies; they serve ``as reference'' when expressed in the
same paradigm as the generated artifact, and ``as inspiration''
otherwise.}

\label{tab:prompts-configurations}
\small
\setlength{\tabcolsep}{4pt}
\begin{tabularx}{\textwidth}{@{}ll>{\raggedright\arraybackslash}X>{\raggedright\arraybackslash}X@{}}
\toprule
\textbf{Configuration} & \textbf{Generated Artifact} & \textbf{Code Context} & \textbf{Input Strategy Examples} \\
\midrule
\j{Py-Base}  & Python player & Python player interface & None \\
\j{Py-SBP-Core} & Python player & Python player interface \& BPpy-based SBP Taki & None \\
\j{Py-SBP-Strategies} & Python player & Python player interface \& BPpy-based SBP Taki & Hand-authored SBP strategy b-threads (as inspiration) \\
\midrule
\j{SBP-Native}  & SBP strategy b-threads & BPpy-based SBP Taki \& BPpy API & Hand-authored SBP strategy b-threads (as reference) \\
\bottomrule
\end{tabularx}
\end{table*}

\subsection{Baselines and Hand-Authored SBP Strategy Refinements}
\label{sec:baselines}

\paragraph{Random Baselines.}
\j{SBP-Random} and \j{Py-Random} implement the same policy: play a legal
card uniformly at random when possible, and otherwise draw.
\j{SBP-Random} is the SBP Taki implementation without strategic
refinements, whereas \j{Py-Random} uses the Python player interface. We
compare them as a validity control
(Section~\ref{sec:baseline}); \j{SBP-Random} is also the fixed opponent
for RQ3.

\paragraph{Hand-Authored SBP Strategy Refinements.}
Since the three refinements of
Section~\ref{sec:hand-authored-sbp-strategies} are designed by the
researchers, their results are not evidence for RQ1--RQ3. They serve as
heuristic baselines and as the strategy examples in the
\j{Py-SBP-Strategies} and \j{SBP-Native} prompts. The blocking b-thread
of \j{Conserve-Super-TAKI} is reused in the composition test
(Section~\ref{sec:executability}).

\subsection{Simulation Protocol}
\label{sec:simulation-protocol}

All strategies are evaluated through repeated Taki simulations. For each comparison, we run 10{,}000 games using sequential random seeds starting from zero. For reproducibility, the seed schedule is fixed across experiments. To control for the observed first-player advantage, starting positions are balanced across each game sequence, so that each player starts approximately half of the games. With 10{,}000 games, 95\% confidence intervals are about $\pm$1 percentage point overall and $\pm$1.4 per starting position. The fixed seeds also allow games to be paired across experiments. Only the relative order of event priorities affects selection. Each game is capped at 1{,}000 player events; the longest game in the four runs of the generated strategies against \j{SBP-Random} had 391 events.

For each run, we record the winner, starting player, number of events,
draw status, deadlock status, and runtime errors. Win rates are computed
over non-error runs; draws and deadlocks are included in this
denominator and reported separately, whereas runtime errors are treated
as incomplete runs and excluded. The specific pairings, including
additional comparisons with hand-authored SBP baselines, are reported in
Section~\ref{sec:results}.

\section{\uppercase{Results}}
\label{sec:results}

We first validate the random baselines and then address RQ1
(Section~\ref{sec:executability}), RQ2
(Section~\ref{sec:sbp-native-results}), and RQ3
(Section~\ref{sec:domain-knowledge-progression}).

\subsection{Random-Baseline Validity Control}
\label{sec:baseline}

We compare \j{SBP-Random} and \j{Py-Random} over 10{,}000 simulated 
games to assess whether the two implementations provide a suitable 
common reference point. As shown in 
Table~\ref{tab:random-baselines}, \j{SBP-Random} wins 50.3\% of the 
games, whereas \j{Py-Random} wins 49.7\%. A two-sided binomial test 
does not indicate a deviation from an even split ($p \approx 0.58$).
No game ended in a draw, deadlock, or runtime error, and games
averaged 93.5 events.

The starting player wins 52.1\% of the games, and both strategies 
perform better when starting than when playing second. Thus, the 
comparison reveals a modest starting-position advantage but no 
evidence of a systematic performance advantage for either random 
implementation. We therefore use \j{SBP-Random} as the common 
baseline in the subsequent evaluations.

\begin{table}[htbp]
\centering
\small
\setlength{\tabcolsep}{3pt}
\caption{Comparison of \j{SBP-Random} and \j{Py-Random} over
10{,}000 games. CI denotes a 95\% Wilson score confidence interval.}
\label{tab:random-baselines}
\begin{tabular}{lcc}
\hline
Metric & \j{SBP-Random} & \j{Py-Random} \\
\hline
Overall win rate & 50.3\% & 49.7\% \\
Win rate when starting & 52.4\% & 51.8\% \\
Win rate when second & 48.2\% & 47.6\% \\
95\% CI when starting & [51.0\%, 53.8\%] & [50.4\%, 53.2\%] \\
95\% CI when second & [46.8\%, 49.6\%] & [46.2\%, 49.0\%] \\
\hline
\end{tabular}
\end{table}

\subsection{Executability and Integration}
\label{sec:executability}

All three generated Python strategies and the \j{SBP-Native} strategy
complete their simulation runs without observed runtime errors,
deadlocks, or draws.

We further evaluate whether \j{SBP-Native} composes with an 
additional hand-authored b-thread that blocks Super TAKI during an 
open regular TAKI sequence. We test four self-play configurations, 
in which the additional b-thread is assigned to neither player, 
either player individually, or both players. All 40{,}000 games 
complete without observed runtime errors, deadlocks, or draws.
The full position-conditioned results are reported in the supplementary material~\cite{\BPTakiCite}.

Together, these results support an affirmative answer to RQ1: the 
generated strategies execute successfully, and \j{SBP-Native} 
integrates with the tested behavioral extension. They do not 
constitute a general guarantee of correctness or conflict-free 
composition beyond the evaluated configurations.

\subsection{Strategic Performance of \j{SBP-Native}}
\label{sec:sbp-native-results}

Table~\ref{tab:sbp-native-vs-python} reports the head-to-head results.
In each comparison, P0 denotes the Python strategy and P1 denotes
\j{SBP-Native}.

\begin{table*}[t]
\centering
\small
\setlength{\tabcolsep}{3pt}
\caption{Head-to-head performance of the Python strategies and
\j{SBP-Native} over 10{,}000 games each.}
\label{tab:sbp-native-vs-python}
\begin{tabular}{lcccccc}
\hline
Configuration & \multicolumn{2}{c}{Overall} &
\multicolumn{2}{c}{Starting} &
\multicolumn{2}{c}{Second} \\
 & P0 & P1 & P0 & P1 & P0 & P1 \\
\hline
\j{Py-Random} vs. \j{SBP-Native}
& 35.9\% & 64.1\% & 38.9\% & 67.1\% & 32.9\% & 61.1\% \\
\j{Py-Base} vs. \j{SBP-Native}
& 47.7\% & 52.3\% & 50.9\% & 55.6\% & 44.4\% & 49.1\% \\
\j{Py-SBP-Core} vs. \j{SBP-Native}
& 48.1\% & 51.9\% & 52.4\% & 56.1\% & 43.9\% & 47.6\% \\
\j{Py-SBP-Strategies} vs. \j{SBP-Native}
& 48.8\% & 51.2\% & 53.5\% & 55.8\% & 44.2\% & 46.5\% \\
\hline
\end{tabular}
\end{table*}

\j{SBP-Native} wins 64.1\% of its games against \j{Py-Random}. Against the three LLM-generated Python strategies, its overall win rate ranges from 51.2\% to 52.3\%. Its win rate is 2.3 percentage points above 50\% against \j{Py-Base}, narrowing to 1.2 percentage points against \j{Py-SBP-Strategies}, which records the strongest result among the generated Python strategies in the preceding baseline comparison. All three head-to-head win rates differ significantly from 50\% ($p < 0.001$ against \j{Py-Base} and \j{Py-SBP-Core}; $p = 0.021$ against \j{Py-SBP-Strategies}).

When playing second, \j{SBP-Native} wins 46.5\%--49.1\% of its games
and the Python strategies 43.9\%--44.4\%; \j{SBP-Native} is higher in
each comparison, although all strategies perform better when starting.

We additionally compare \j{SBP-Native} against three hand-authored SBP
strategy refinements. Its overall win rate ranges from 52.3\% to 52.5\%
in these matchups; the full results appear in the accompanying
supplementary material.

Together, these results support an affirmative answer to RQ2 within the
evaluated configurations: the directly generated \j{SBP-Native}
strategy achieves competitive performance relative to all 
generated Python strategies. The observed margins remain small,
particularly against \j{Py-SBP-Strategies}, and therefore do not
support a general claim that SBP strategies are superior to 
Python ones.

\subsection{Effect of Domain Knowledge Cast through SBP}
\label{sec:domain-knowledge-progression}

Table~\ref{tab:domain-knowledge-progression} reports the performance of
the generated strategies against the same \j{SBP-Random} baseline.

\begin{table*}[t]
\centering
\small
\setlength{\tabcolsep}{3pt}
\caption{Performance against \j{SBP-Random} over 10{,}000 games each.
See Sections~\ref{sec:executability} and~\ref{sec:sbp-native-results}
for the composition and head-to-head results, respectively.}
\label{tab:domain-knowledge-progression}
\begin{tabular}{lccccc}
\hline
Strategy & Overall & Starting & Second & 95\% CI starting & 95\% CI second \\
\hline
\j{Py-Base} & 59.4\% & 62.3\% & 56.5\% & [61.0\%, 63.7\%] & [55.1\%, 57.9\%] \\
\j{Py-SBP-Core} & 59.8\% & 63.3\% & 56.4\% & [61.9\%, 64.6\%] & [55.0\%, 57.7\%] \\
\j{Py-SBP-Strategies} & 62.5\% & 65.7\% & 59.2\% & [64.4\%, 67.0\%] & [57.9\%, 60.6\%] \\
\hline
\j{SBP-Native} & 64.9\% & 67.4\% & 62.4\% & [66.1\%, 68.7\%] & [61.0\%, 63.7\%] \\
\hline
\end{tabular}
\end{table*}

All three generated Python strategies outperform the random baseline,
with overall win rates ranging from 59.4\% to 62.5\%. The difference
between \j{Py-Base} and \j{Py-SBP-Core} is not significant
($p = 0.42$), whereas \j{Py-SBP-Strategies} differs significantly
from both ($p < 0.001$). The progression is not monotonic when
playing second.

\j{Py-SBP-Strategies} records the highest win rate overall and in both
turn positions; overall, it is 3.1 percentage points above
\j{Py-Base} and 2.7 above \j{Py-SBP-Core}. The separately evaluated
\j{SBP-Native} strategy wins 64.9\% of games against
\j{SBP-Random}, so it also outperforms this baseline.

Together, these results support an affirmative answer to RQ3 for the
evaluated configurations. Richer domain knowledge cast through SBP is
associated with stronger generated Python strategies, with the highest
observed Python performance obtained when example strategy b-threads
are included. Because we evaluate one generated artifact per
configuration, these results indicate an association and do not isolate
a causal effect of any individual context component.

\section{\uppercase{Discussion}}
\label{sec:discussion}

\noindent\textbf{Findings and Implications.} The Python progression suggests that SBP-cast domain knowledge can help guide LLM-generated strategies, although its effect was uneven: adding the SBP implementation alone changed the win rate little (59.4\% to 59.8\%), whereas the highest rate (62.5\%) was obtained when example strategy b-threads were also provided. While the SBP implementation already specifies, through blocking, which actions are illegal, the examples additionally encode strategic choices, such as prioritizing TAKI-family cards or withholding an otherwise legal Super TAKI; we did not test whether this accounts for the difference. With one artifact per configuration, we cannot isolate the effect of each context component.

The \j{SBP-Native} configuration evaluates direct generation of strategy b-threads: in our simulations, the generated b-threads execute without observed runtime errors, deadlocks, or other failures, and compose with an additional hand-authored b-thread, suggesting SBP can provide a useful integration boundary for LLM-generated reactive code. The comparison between Python and SBP-native strategies also indicates that the target representation affects performance: Python strategies implement centralized heuristics, whereas SBP-native strategies influence event selection through modular b-threads that need not reproduce existing rule logic. Since \j{SBP-Native} is only slightly ahead of \j{Py-SBP-Strategies}, we interpret this as evidence of competitiveness rather than broad superiority.

\medskip
\noindent\textbf{Threats to Validity.} Several threats limit the generality of these findings. First, we study a single domain. Taki is a two-player, turn-based game, and our implementation covers a subset of its rules. The observed results may not hold for systems with asynchronous inputs, timing constraints, concurrent physical processes, or safety-critical requirements. Second, we rely on a single SBP framework, BPpy, and one Taki implementation built with it. The results may depend on its API, its priority-based event selection, and the way our b-threads decompose the game; we do not know whether they carry over to other SBP frameworks or implementations. Third, we evaluate one generated artifact per configuration, produced with one target model and a specific set of prompts, so we cannot measure how results vary across generations, models, or prompts. Relatedly, our significance tests compare specific generated strategies across game seeds; they do not show that a configuration reliably yields stronger strategies. Statistical significance also does not imply practical importance: the head-to-head margins are only 1.2--2.3 percentage points. Finally, the evaluation is simulation-based, so the absence of observed runtime errors, deadlocks, or composition failures is not a formal correctness guarantee; and win rate captures strategic effectiveness but not code quality, maintainability, comprehensibility, or ease of human review.

\medskip
\noindent\textbf{Applicability to Other Systems.} The approach is not specific to the rules of Taki, but applying it elsewhere presupposes that relevant system behavior can be observed and constrained through events, that existing behavior and domain knowledge can be expressed as b-threads, and that generated extensions can be evaluated against suitable tests or simulations. SBP has been used in settings such as IoT and satellite software~\cite{ElMaStWe18,BaElSaWe19}, which are natural candidates for further study. These earlier uses of SBP do not show that our LLM-related results transfer to such domains. Systems with timing, interrupt, or safety requirements would also need evaluation methods that address those properties directly.

\section{\uppercase{Related Work}}
\label{sec:related-work}

Large language models are increasingly used for software-development tasks such as code generation, debugging, and program repair, with early work showing both their ability to synthesize programs from natural-language descriptions and the challenges in producing functionally correct code~\cite{ChTwJuYuDeKaEdBuJoBr21,LiXiWaZh23}. Recent coding agents and repository-level benchmarks highlight the importance of repository context, tool use, and cross-file reasoning for realistic tasks~\cite{Op25,JiYaWeYaPePrNa24,LiXuMc24}. This is especially relevant for reactive systems, where correctness depends on coordinated behavior over time; recent work has explored LLMs for reactive synthesis and modeling support~\cite{MuHoQiCuRoKoSa24,HaMa24}.

Recent work also combines LLMs with scenario-based programming.
One approach uses SBP as a design structure for
LLM-assisted development, translating requirements directly into
b-threads composed by the SBP execution mechanism~\cite{YaElWe24}.
Another integrates LLMs into scenario-based modeling, combining
generation with manual inspection, simulation, and formal
analysis~\cite{HaKaMaSz25}. A related SBP-based workflow for Connect4
decomposes requirements into scenarios generated one b-thread at a time
and refined via inspection and, when applicable, verifier
feedback~\cite{BeKa26}.

SBP has been shown to naturally lend itself to various
verification-based tasks, such as repair~\cite{HaKaMaWe14,HaKaMaWe12},
optimization~\cite{HaKaKa13}, and compositional
analysis~\cite{HaKaKaMaMiWe13}. It will be interesting to explore the
synergies between our approach and these capabilities.

Unlike this prior work, which primarily focuses on generating scenario-based models or requirements from natural-language specifications, we study LLM-generated extensions to an existing reactive system---shifting the challenge from generating standalone behaviors to composing new code with existing components.

\section{\uppercase{Conclusions and Future Work}}
\label{sec:conclusions}

Using the Taki card game as a case study, we find that
LLM-generated SBP strategy b-threads integrate into the existing
program without observed runtime errors, deadlocks, or draws in the
evaluated simulations, while achieving competitive performance
relative to LLM-generated Python strategies. In addition, richer
domain knowledge cast through SBP is associated with stronger
generated Python strategies in our simulations. These findings
suggest that SBP can provide a useful structured representation for
both communicating domain knowledge and generating composable
extensions for reactive software.

More broadly, our study highlights  SBP's potential  as a
knowledge representation medium for cooperation between software
engineers and AI models during software development.
In our experiments, an SBP implementation and example b-threads were
used to provide domain knowledge to the model, and in the
\j{SBP-Native} configuration the model expressed its contribution as
b-threads that composed with the existing program. Whether this
representation also helps engineers understand, review, and refine
such contributions requires studies with software engineers, which we
have not conducted.

In future work, we plan to apply the proposed approach to reactive systems in other application domains, such as IoT and autonomous robotic systems, to examine which of our findings carry over. We will also evaluate additional SBP frameworks and implementations, LLMs, and prompting strategies, as well as study software-engineering aspects beyond gameplay performance, including code quality, readability, maintainability, comprehensibility, and human review effort. In particular, we plan to study how the approach scales to more complex domains, with larger sets of interacting b-threads and events, longer model contexts, and asynchronous or timing-dependent behavior. Another promising direction is to investigate richer forms of LLM-assisted SBP, such as composing multiple generated b-threads or iteratively refining them based on simulation or formal-analysis feedback.

\section*{\uppercase{Acknowledgements}}

\noindent\textbf{AI Tools:} \emph{Claude Sonnet 4.6}~\cite{An26} assisted with manuscript editing and \LaTeX{} formatting. \emph{Codex}~\cite{Op25} and \emph{GPT-5.4}~\cite{Op26} were used as described in Section~\ref{sec:experimental-pipeline}.

\ifblindreview
\else

This research was partially funded by the European Union, as part of the RobustifAI project (ID 101212818). Views and opinions expressed are however those of the author(s) only and do not necessarily reflect those of the European Union or the European Health and Digital Executive Agency (HADEA). Neither the European Union nor the granting authority can be held responsible for them. In addition, this work was partially supported
by the Israeli Smart Transportation Research Center (ISTRC). The work of Tom Yaacov is supported by the Causality in Healthcare AI (CHAI) Hub [UKRI AI and EPSRC grant EP/Y028856/1].
\fi

\bibliographystyle{apalike}
{\small
\bibliography{bibliography}}

@inproceedings{YaWeAsKaZi25,
  title={{Exploring and Evaluating Interplays of BPpy with Deep Reinforcement Learning and Formal Methods}},
  author={Yaacov, Tom and Weiss, Gera and Ashrov, Adiel and Katz, Guy and Zisser, Jules},
  booktitle={Proc. 20th Int. Conf. on Evaluation of Novel Approaches to Software Engineering (ENASE)},
  year={2025}
}

@inproceedings{GoMaMe12,
  title={{Spaghetti for the Main Course? Observations on the Naturalness of Scenario-Based Programming}},
  author={M. Gordon and A. Marron and O. Meerbaum-Salant},
  booktitle={Proc. 17th Conf. on Innovation and Technology in Computer Science Education (ITiCSE)},
  year={2012}
}

@misc{MuHoQiCuRoKoSa24,
  title={{Combining LLM Code Generation with Formal Specifications and Reactive Program Synthesis}},
  author={Murphy, William and Holzer, Nikolaus and Qiao, Feitong and Cui, Leyi and Rothkopf, Raven and Koenig, Nathan and Santolucito, Mark},
  note={Technical Report. \url{http://arxiv.org/abs/2410.19736}},
  year={2024}
}

@misc{HaMa24,
  title={{Preparing for Super-Reactivity: Early Fault-Detection in the Development of Exceedingly Complex Reactive Systems}},
  author={Harel, David and Marron, Assaf},
  note={Technical Report. \url{http://arxiv.org/abs/2410.02627}},
  year={2024}
}

@book{HaMa03,
  title={{Come, Let's Play: Scenario-Based Programming Using LSCs and the Play-Engine}},
  author={Harel, David and Marelly, Rami},
  publisher={Springer},
  year={2003}
}

@inproceedings{LiXuMc24,
  title={{RepoBench: Benchmarking Repository-Level Code Auto-Completion Systems}},
  author={Liu, Tianyang and Xu, Canwen and McAuley, Julian},
  booktitle={Proc. 12th Int. Conf. on Learning Representations (ICLR)},
  year={2024}
}

@inproceedings{JiYaWeYaPePrNa24,
  title={{SWE-bench: Can Language Models Resolve Real-World GitHub Issues?}},
  author={Jimenez, Carlos E. and Yang, John and Wettig, Alexander and Yao, Shunyu and Pei, Kexin and Press, Ofir and Narasimhan, Karthik},
  booktitle={Proc. 12th Int. Conf. on Learning Representations (ICLR)},
  year={2024}
}

@misc{ShGa26b,
  author={{Shafir Games}},
  title={{TAKI Video Tutorial}},
  note={\url{https://youtu.be/5-pFhiwwAME}},
  year={2026}
}

@misc{ShGa26a,
  author={{Shafir Games}},
  title={{TAKI Card Game}},
  note={\url{https://www.takigame.com/}},
  year={2026}
}

@misc{AsYaKa26a,
  author={Ashrov, Adiel and Yaacov, Tom and Katz, Guy},
  title={{BP-Taki: A Behavioral Programming Implementation of the Taki Card Game}},
  note={\url{https://github.com/adielashrov/bp-taki}},
  year={2026}
}

@misc{An26,
  author={{Anthropic}},
  title={{Claude Sonnet 4.6 System Card}},
  note={\url{https://www-cdn.anthropic.com/78073f739564e986ff3e28522761a7a0b4484f84.pdf}},
  year={2026}
}

@misc{Op25,
  author={{OpenAI}},
  title={{Codex}},
  note={\url{https://openai.com/codex/}},
  year={2025}
}

@misc{Op26,
  author={{OpenAI}},
  title={{Introducing GPT-5.4}},
  note={\url{https://openai.com/index/introducing-gpt-5-4/}},
  year={2026}
}

@inproceedings{HaMaWe10,
  title={{Programming Coordinated Behavior in Java}},
  author={Harel, David and Marron, Assaf and Weiss, Gera},
  booktitle={Proc. 24th European Conf. on Object-Oriented Programming (ECOOP)},
  pages={250--274},
  year={2010}
}

@inproceedings{BaWeSh18,
  title={{BPjs: An Extensible, Open Infrastructure for Behavioral Programming Research}},
  author={Bar-Sinai, Michael and Weiss, Gera and Shmuel, Reut},
  booktitle={Proc. 21st ACM/IEEE Int. Conf. on Model Driven Engineering Languages and Systems: Companion Proceedings (MODELS)},
  year={2018}
}

@inproceedings{HaKa14,
  title={{Scaling-Up Behavioral Programming: Steps from Basic Principles to Application Architectures}},
  author={Harel, David and Katz, Guy},
  booktitle={Proc. 4th Int. Workshop on Programming Based on Actors, Agents and Decentralized Control (AGERE!)},
  pages={95--108},
  year={2014}
}

@inproceedings{ElMaStWe18,
  title={{A Context-Based Behavioral Language for IoT}},
  author={Elyasaf, Achiya and Marron, Assaf and Sturm, Arnon and Weiss, Gera},
  booktitle={Proc. 21st Int. Conf. on Model Driven Engineering Languages and Systems Workshops (MODELS)},
  year={2018}
}

@inproceedings{BaElSaWe19,
  title={{A Scenario Based On-Board Software and Testing Environment for Satellites}},
  author={Bar-Sinai, M and Elyasaf, A and Sadon, A and Weiss, G},
  booktitle={Proc. 59th Israel Annual Conf. on Aerospace Sciences (IACAS)},
  year={2019}
}

@article{HaMaWe12,
  title={{Behavioral Programming}},
  author={Harel, David and Marron, Assaf and Weiss, Gera},
  journal={Communications of the ACM},
  volume={55},
  number={7},
  pages={90--100},
  year={2012}
}

@misc{ChTwJuYuDeKaEdBuJoBr21,
  title={{Evaluating Large Language Models Trained on Code}},
  author={Chen, Mark and Tworek, Jerry and Jun, Heewoo and Yuan, Qiming and de Oliveira Pinto, Henrique Ponde and Kaplan, Jared and Edwards, Harrison and Burda, Yuri and Joseph, Nicholas and Brockman, Greg and others},
  note={Technical Report. \url{http://arxiv.org/abs/2107.03374}},
  year={2021}
}

@inproceedings{LiXiWaZh23,
  title={{Is Your Code Generated by ChatGPT Really Correct? Rigorous Evaluation of Large Language Models for Code Generation}},
  author={Liu, Jiawei and Xia, Chunqiu Steven and Wang, Yuyao and Zhang, Lingming},
  booktitle={Proc. 37th Annual Conf. on Neural Information Processing Systems (NeurIPS)},
  year={2023}
}

@inproceedings{YaElWe24,
  title={{Boosting LLM-Based Software Generation by Aligning Code with Requirements}},
  author={Yaacov, Tom and Elyasaf, Achiya and Weiss, Gera},
  booktitle={Proc. 32nd IEEE Int. Requirements Engineering Conf. Workshops (REW)},
  year={2024}
}

@inproceedings{HaKaMaSz25,
  title={{Enhancing Scenario-Based Modeling Using Large Language Models}},
  author={Harel, David and Katz, Guy and Marron, Assaf and Szekely, Smadar},
  booktitle={Proc. 12th Int. Conf. on Model-Driven Engineering and Software Development (MODELSWARD)},
  pages={43--68},
  year={2025}
}

@inproceedings{BeKa26,
  title={{On Integrating Large Language Models and Scenario-Based Modeling for Improving Software Reliability}},
  author={Berzack, Ayelet and Katz, Guy},
  booktitle={Proc. 14th Int. Conf. on Model-Based Software and Systems Engineering (MODELSWARD)},
  pages={21--32},
  year={2026}
}

@article{DaHa01,
  title={{LSCs: Breathing Life into Message Sequence Charts}},
  author={Damm, Werner and Harel, David},
  journal={Formal Methods in System Design},
  volume={19},
  number={1},
  pages={45--80},
  year={2001}
}

@inproceedings{LiLiZhLiLiZhWaLiFaWaDiZhZhDoJiLiHuLiGuYa24,
  title={{DevEval: A Manually-Annotated Code Generation Benchmark Aligned with Real-World Code Repositories}},
  author={Li, Jia and Li, Ge and Zhao, Yunfei and Li, Yongmin and Liu, Huanyu and Zhu, Hao and Wang, Lecheng and Liu, Kaibo and Fang, Zheng and Wang, Lanshen and Ding, Jiazheng and Zhang, Xuanming and Zhu, Yuqi and Dong, Yihong and Jin, Zhi and Li, Binhua and Huang, Fei and Li, Yongbin and Gu, Bin and Yang, Mengfei},
  booktitle={Proc. 62nd Annual Meeting of the Association for Computational Linguistics (ACL)},
  pages={3603--3614},
  year={2024}
}

@inproceedings{HaNgBu25,
  title={{On the Impacts of Contexts on Repository-Level Code Generation}},
  author={Hai, Nam Le and Nguyen, Dung Manh and Bui, Nghi D. Q.},
  booktitle={Proc. Conf. Nations of the Americas Chapter of the Association for Computational Linguistics (NAACL)},
  pages={1496--1524},
  year={2025}
}

@article{Mc47,
  title={{Note on the Sampling Error of the Difference Between Correlated Proportions or Percentages}},
  author={McNemar, Quinn},
  journal={Psychometrika},
  volume={12},
  number={2},
  pages={153--157},
  year={1947}
}

@article{Ho79,
  title={{A Simple Sequentially Rejective Multiple Test Procedure}},
  author={Holm, Sture},
  journal={Scandinavian Journal of Statistics},
  volume={6},
  number={2},
  pages={65--70},
  year={1979}
}

@inproceedings{GrGrKaMa16,
 author = {Greenyer, J. and Gritzner, D. and Katz, G. and Marron, A.},
 title = {{Scenario-Based Modeling and Synthesis for Reactive Systems with Dynamic System Structure in ScenarioTools}},
 booktitle = {Proc. 19th ACM/IEEE Int. Conf. on Model Driven Engineering Languages and Systems (MODELS)},
 year = {2016},
 pages = {16--23},
}

@article{HaKaMaWe14,
 author = {Harel, D. and Katz, G. and Marron, A. and Weiss, G.},
 title = {{Non-Intrusive Repair of Safety and Liveness Violations in Reactive Programs}},
 journal = {Transactions on Computational Collective Intelligence (TCCI)},
 year = {2014},
 volume = {16},
 pages = {1--33},
}

@inproceedings{HaKaKa13,
title = {{Relaxing Synchronization Constraints in Behavioral Programs}},
author = {Harel, D. and Kantor, A. and Katz, G.},
booktitle = {Proc. 19th Int. Conf. on Logic for Programming, Artificial Intelligence and Reasoning (LPAR)},
year = {2013},
}

@inproceedings{HaKaKaMaMiWe13,
author = {Harel, D. and Kantor, A. and Katz, G. and Marron, A. and Mizrahi, L. and Weiss, G.},
title = {{On Composing and Proving the Correctness of Reactive Behavior}},
booktitle = {Proc. 13th Int. Conf. on Embedded Software (EMSOFT)},
year = {2013},
}

@inproceedings{HaKaMaWe12,
 author = {Harel, D. and Katz, G. and Marron, A. and Weiss, G.},
 title = {{Non-Intrusive Repair of Reactive Programs}},
 booktitle = {Proc. 17th IEEE Int. Conf. on Engineering of Complex Computer Systems (ICECCS)},
 year = {2012},
 pages = {3--12},
}

\end{document}